\documentclass[12pt]{article}
\usepackage[mathscr]{eucal}
\usepackage{amssymb,latexsym}
\usepackage{verbatim}
\usepackage{amsmath}
\usepackage{amsthm}
\usepackage{enumerate}
\usepackage{authblk}
\usepackage{color}
\usepackage{url}

\usepackage{tikz}
\usetikzlibrary{matrix}
\usetikzlibrary{snakes}
\usetikzlibrary{arrows,calc,shapes,decorations.pathreplacing}
\usepackage{tikz-cd}

\usepackage[normalem]{ulem}

\usepackage{amscd,amssymb,amsthm,verbatim}
\usepackage{enumerate}
\usepackage{bm}
\usepackage{mathrsfs, graphicx} 
\usepackage{stmaryrd}

\usepackage[nottoc,notlot,notlof]{tocbibind}
\usepackage{setspace}

\usepackage[%
bookmarks=true,
colorlinks,
linkcolor=blue,
urlcolor=blue,
citecolor=blue,
plainpages=false,
pdfpagelabels,
final,
breaklinks=true\between
]{hyperref}
\usepackage{hyperref}

\newtheorem{thm}{Theorem}
\newtheorem{lem}[thm]{Lemma}
\newtheorem{prop}[thm]{Proposition}

\renewcommand\S{\Sigma}

\newcommand\D{\nabla}

\newcommand\ric{{\rm Ric}}

\newcommand\beq{\begin{equation}}
\newcommand\eeq{\end{equation}}
\newcommand\ben{\begin{enumerate}}
\newcommand\een{\end{enumerate}}
\newcommand\bit{\begin{itemize}}
\newcommand\eit{\end{itemize}}

\newcounter{mnotecount}

\title{A simplified proof of a cosmological singularity theorem}

\author[1]{Gregory J. Galloway\footnote{galloway@math.miami.edu}}
\author[2]{Eric Ling\footnote{el@math.ku.dk}}

\affil[1]{University of Miami, Coral Gables, FL, USA}

\affil[2]{Copenhagen Centre for Geometry and Topology (GeoTop),
\linebreak
Department of Mathematical Sciences, \linebreak University of Copenhagen, Denmark}

\begin{document}
\date{}
\maketitle

\vspace{.15in}

\begin{abstract} 
In a previous paper \cite{GL_topo_cosmo}, we proved the following singularity theorem applicable to cosmological models with a positive cosmological constant: if a four-dimensional spacetime satisfying the null energy condition contains a compact Cauchy surface which is expanding in all directions, then the spacetime is past null geodesically incomplete unless the Cauchy surface is topologically a spherical space. The proof in \cite{GL_topo_cosmo} made use of the positive resolution of the surface subgroup conjecture \cite{Kahn}. In this note, we demonstrate how the less-broadly-known positive resolution of the virtual positive first Betti number conjecture \cite{Agol} provides a more streamlined and unified approach to the proof. We illustrate the theorem with some examples and analyze its rigidity under null geodesic completeness.
\end{abstract}


\section{Introduction}

A theme of long-standing interest in general relativity  concerns the relationship between the topology of spacetime and the occurrence of singularities, by which we mean here 
causal geodesic incompleteness.  Many such results center around the notion of {\it topological censorship}; see e.g.,\ \cite{Gannon, Lee, GLCut, FSW, GalTopo, GSWW, EGP}, which concerns the topology of the domain of outer communications in black hole spacetimes.

In \cite{GL_topo_cosmo}, we presented a result relating topology and singularities in the {\it cosmological setting}, by which in general we mean  globally hyperbolic spacetimes with compact Cauchy surfaces.  Let us briefly recall the motivation for this result.

Hawking's classical cosmological singularity theorem \cite[p. 272]{HE} establishes past timelike geodesic incompleteness in spacetimes $(M,g)$  that admit  compact spacelike hypersurfaces $V$  that are {\it future expanding}, i.e, which have positive mean curvature, 
${\rm tr} K > 0$. (Here $K$ is the second fundamental form of $V$: For vectors $X,Y \in T_pV$, $K(X,Y) = g(\D_Xu, Y)$ and $u$ is the future directed unit timelike normal to $V$.)
%
%

Hawking's theorem requires the Ricci tensor of the spacetime to satisfy the strong energy condition (SEC),  $\ric(X,X) \ge 0$ for all timelike vectors~$X$.  However, solutions to the Einstein equations with positive cosmological constant,  $\Lambda > 0$, will not in general satisfy the strong energy condition, and the conclusion to Hawking's theorem will not in general hold -- de Sitter space  is a prime example.  It is geodesically complete and (in $3+1$ dimensions) satisfies 
$$
\ric = \Lambda g \qquad (\Lambda > 0)  \,.
$$
While de Sitter space fails to satisfy the SEC, it does satisfy the null energy condition (NEC), $\ric(X,X) \geq 0$ for all null vectors $X$.  In fact, under a mild condition on the energy-momentum tensor, any spacetime $(M,g)$, satisfying the Einstein equations with cosmological constant, satisfies the NEC.

%


In \cite{GL_topo_cosmo} we proved the following theorem.
\begin{thm}\label{thm: main}
Let $(M,g)$ be a four-dimensional spacetime satisfying the null energy condition. 
 Assume there is a smooth spacelike compact Cauchy surface $V$ which is expanding in all directions (i.e., its second fundamental form $K$ is positive definite). Then either 
\begin{itemize}
\item[(i)] $V$ is a spherical space (i.e., it is topologically a quotient of $S^3$), or
\item[(ii)] $M$ is past null geodesically incomplete. 
\end{itemize} 
\end{thm}

This theorem may be viewed as a cosmological singularity theorem for spacetimes compatible with a positive cosmological constant:  If the Cauchy surface has a topology other than that of a spherical space, the spacetime is past null geodesically incomplete.  De Sitter space, and certain quotients, illustrate the relevance of case~(i).  
Dust-filled FLRW models, with positive cosmological constant and with compact Cauchy surfaces of positive, negative, and zero curvature illustrate both cases. The negative and zero curvature cases do not have spherical space topology and all begin with a big bang. For a suitably chosen value of the cosmological constant, the positive curvature (i.e., spherical) case can escape a big bang singularity. See e.g.,\ 
\cite[Section~23.3]{DInverno}.

The following construction yields a class of $\Lambda$-vacuum examples of Theorem \ref{thm: main}.

\medskip

\noindent
\emph{Further examples.} Let $V$ be a compact 3-manifold. By the Yamabe problem (see \cite{Schoen1989} for a discussion), there is a Riemannian metric $h$ on $V$ with constant scalar curvature $R_h$. The $\Lambda$-vacuum Einstein constraint equations are
\[
R_h - |K|_h^2 + (\text{tr}_hK)^2 = 2\Lambda \quad \text{ and } \quad D_iK^i_{\:j} - D_jK^i_{\:i} = 0,
\]
{where $D$ is the $h$-covariant derivative. Set $K = h$. Then the second constraint is satisfied since $DK = Dh = 0$. The left-hand side of the first constraint is constant; choose $\Lambda$ so that it's satisfied. Since $K$ is positive definite, Theorem \ref{thm: main} implies that the resulting maximal globally hyperbolic development of the initial data set $(V, h, K=h)$ is past null geodesically incomplete so long as $V$ is not a spherical space.

\medskip

The proof of Theorem \ref{thm: main} in \cite{GL_topo_cosmo} made use of a deep consequence of Thurston's geometrization conjecture,  namely the positive resolution of the surface subgroup conjecture \cite{Kahn}.  In this paper, we show, alternatively, how the less-broadly-known positive resolution of the virtual positive first Betti number conjecture \cite{Agol} (another deep consequence of Thurston's geometrization conjecture), can be used to  simplify  aspects of the proof, which thereby provides a unified approach to the proof.

Our simplified proof of Theorem 1 is presented in the next section.  The proof is ultimately an application of Penrose's singularity theorem \cite{Penrose} (applied to a {\it past trapped} surface).  Recently, we studied certain {\it rigidity} aspects of Penrose's singularity theorem~\cite{GL_Penrose_rigid}.  In Section 3 we present a `rigid version' of Theorem \ref{thm: main},  whose proof makes use of a result in \cite{GL_Penrose_rigid}.


\section{The simplified proof of Theorem \ref{thm: main}}

We will make use of the following lemma and proposition. They appear as Lemma 4 and Proposition 5 in \cite{GL_topo_cosmo}. 

\begin{lem}\label{lem}
Let $(M,g)$ be a spacetime with a smooth spacelike Cauchy surface $V$. Suppose $p \colon \tilde{V} \to V$ is a Riemannian covering. Then there exists a Lorentzian covering $P \colon \tilde{M} \to M$ such that $\tilde{V}$ is a Cauchy surface for $\tilde{M}$ and $P|_{\tilde{V}} = p$. 
\end{lem}

This lemma can be proved using  basic covering space theory (cf. beginning of section 3.2 in \cite{GLCut}).  Also, in \cite{GL_topo_cosmo} a somewhat more direct proof is given using a splitting result of Bernal and S\'anchez~\cite{Sanchez}.

In what follows we will always use $\mathbb{Z}$ coefficients for homology.}

\begin{prop}\label{prop}
Let $(M,g)$ be a four-dimensional spacetime satisfying the null energy condition. Let $V$ be an oriented smooth compact spacelike Cauchy surface that is expanding in all directions. If $V$ has nontrivial second homology, $H_2(V) \neq 0$, then $M$ is past null geodesically incomplete.
\end{prop}

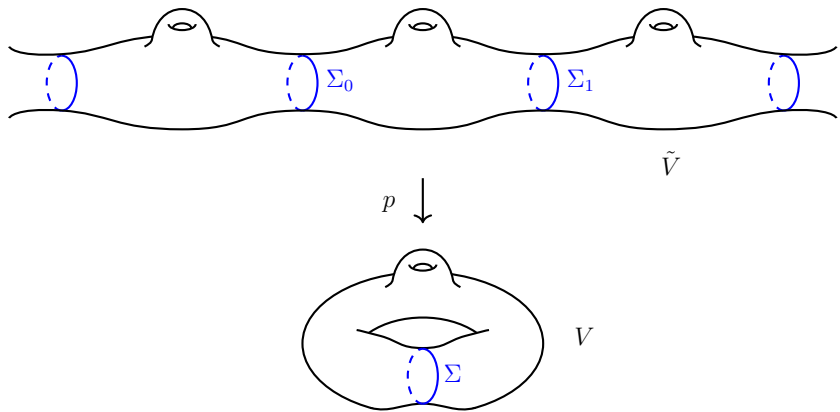
\begin{figure}
\[
\begin{tikzpicture}
\draw[thick] (0,-0.1) ellipse (1.6 and .95);
\begin{scope}[scale=.8]
\draw [thick] plot [smooth,tension=1] coordinates {(-1.1,.1) (-0.7,0) (0,-.2) (0.7,0) (1.1,.1)};
\draw[rounded corners=24pt,thick] (-.9,0.05)--(0,.65)--(.9,0.05);
\end{scope}
\draw[thick,white,fill=white] (0,-1) ellipse (.7 and 0.1);
\draw [thick] (-.7,-.955)  to[out=-15,in=180]  (0,-.9) to[out=0,in=195]  (.7,-.955);


\draw[dashed,thick,blue] (0,-.9) arc (270:90:.2 and .365);16\draw[thick,blue]  (0,-.9) arc (-90:90:.2 and .365);
\draw (0.4, -.5) node [scale=.8,blue]{$\S$};

\draw (2.15, 0) node [scale=.8]{$V$};

\draw[thick,white,fill=white] (0,.8) ellipse (.37 and 0.1);
\draw [thick] (0.18,.95) arc(0:-180: 0.18 and .08);
\draw [thick] (0.12,0.9) arc(0:180: 0.12 and .06);
\draw [thick] (-0.5,0.65) to[out=30,in=-100] (-0.4,0.75) to[out=80,in=180]   (0,1.15) to[out=0,in=100]  (0.4,0.75) to[out=-80,in=150] (0.5,0.65);

\draw [->,thick] (0,2.1) -- (0,1.5);
\draw (-.45, 1.75) node [scale=.8]{$p$};

\draw[dashed,thick,blue] (-1.6,3) arc (270:90:.2 and .365);
\draw[thick,blue]  (-1.6,3) arc (-90:90:.2 and .365);
\draw (-1.1, 3.4) node [scale=.8,blue]{$\S_0$};
\draw[dashed,thick,blue] (1.6,3) arc (270:90:.2 and .365);
\draw[thick,blue]  (1.6,3) arc (-90:90:.2 and .365);
\draw (2.1, 3.4) node [scale=.8,blue]{$\S_1$};
\draw[dashed,thick,blue] (4.8,3) arc (270:90:.2 and .365);
\draw[thick,blue]  (4.8,3) arc (-90:90:.2 and .365);
\draw[dashed,thick,blue] (-4.8,3) arc (270:90:.2 and .365);
\draw[thick,blue]  (-4.8,3) arc (-90:90:.2 and .365);

\draw [thick] plot [smooth, tension=1] coordinates {(-5.5,2.9) (-4.8,3) (-3.2,2.75) (-1.6,3) (0,2.75) (1.6,3) (3.2,2.75) (4.8,3) (5.5,2.9) };
\draw [thick] plot [smooth, tension=1] coordinates {(-5.5,3.85) (-4.8,3.75) (-3.2,4) (-1.6,3.75) (0,4) (1.6,3.75) (3.2,4) (4.8,3.75) (5.5,3.85) };

\draw (3.3, 2.3) node [scale=.8]{$\tilde{V}$};

\begin{scope}[shift = {(0,3.2)}]
\draw[thick,white,fill=white] (0,.8) ellipse (.38 and 0.15);
\draw [thick] (0.18,.95) arc(0:-180: 0.18 and .08);
\draw [thick] (0.12,0.9) arc(0:180: 0.12 and .06);
\draw [thick] (-0.5,0.65) to[out=30,in=-100] (-0.4,0.75) to[out=80,in=180]   (0,1.15) to[out=0,in=100]  (0.4,0.75) to[out=-80,in=150] (0.5,0.65);
\end{scope}

\begin{scope}[shift = {(3.2,3.2)}]
\draw[thick,white,fill=white] (0,.8) ellipse (.38 and 0.15);
\draw [thick] (0.18,.95) arc(0:-180: 0.18 and .08);
\draw [thick] (0.12,0.9) arc(0:180: 0.12 and .06);
\draw [thick] (-0.5,0.65) to[out=30,in=-100] (-0.4,0.75) to[out=80,in=180]   (0,1.15) to[out=0,in=100]  (0.4,0.75) to[out=-80,in=150] (0.5,0.65);
\end{scope}

\begin{scope}[shift = {(-3.2,3.2)}]
\draw[thick,white,fill=white] (0,.8) ellipse (.38 and 0.15);
\draw [thick] (0.18,.95) arc(0:-180: 0.18 and .08);
\draw [thick] (0.12,0.9) arc(0:180: 0.12 and .06);
\draw [thick] (-0.5,0.65) to[out=30,in=-100] (-0.4,0.75) to[out=80,in=180]   (0,1.15) to[out=0,in=100]  (0.4,0.75) to[out=-80,in=150] (0.5,0.65);
\end{scope}

\end{tikzpicture}
\]
\caption{The noncompact cover $\tilde{V}$ of $V$ constructed in the proof of Proposition \ref{prop}.}
\label{fig}
\end{figure}

\noindent \emph{Sketch of proof.} By well-known arguments in geometric measure theory, each nontrivial class of $H_2(V)$ has a least area representative which can be expressed as a linear combination of smooth, oriented, connected, compact, embedded minimal surfaces in $V$. Let $\Sigma$ be such a surface. The past null expansions of $\Sigma$ are given by $\theta^{\pm} = -\text{tr}_\Sigma K \pm H$, where $K$ is the second fundamental of $V$ within $M$. Since $V$ is expanding in all directions and $H = 0$ on $\Sigma$, we have $\theta^{\pm} < 0$. Hence $\Sigma$ is past trapped. Let $W$ be the compact manifold with two boundary components formed by cutting $V$ along $\Sigma$. We can pass to a Riemannian covering $p \colon \tilde{V} \to V$ which unravels $V$ by gluing $\mathbb{Z}$ copies of $W$ end-to-end. See Figure \ref{fig}. The cover $\tilde{V}$ is noncompact with $\mathbb{Z}$ isometric copies of the past trapped surface $\Sigma$; hence Penrose's singularity theorem \cite{Penrose} implies that the spacetime $\tilde{M}$ obtained from Lemma \ref{lem} is past null geodesically incomplete. Therefore $M$ is past null geodesically incomplete \cite[Cor. 7.29]{ON}.
\qed

\medskip

\noindent\emph{Proof of Theorem \ref{thm: main}.} Either $V$ has finite fundamental group or infinite fundamental group. In the former case, it follows from the positive resolution of the elliptization conjecture (see \cite{3_mfld_groups} for a discussion) that $V$ is a spherical space. 

Assume that $V$ has infinite fundamental group. We will show that $V$ admits an oriented finite cover $\tilde{V}$ with $H_2(\tilde{V}) \neq 0$. Then past null geodesic incompleteness of $(M,g)$ follows by Proposition \ref{prop}. 

We can assume $V$ is oriented by passing to its oriented double cover. By the prime decomposition theorem of oriented 3-manifolds (see \cite{Hatcher3MOverview} for a discussion), $V$ decomposes uniquely as a connected sum of oriented prime 3-manifolds, 
\[
V = V_1 \# \dotsb \# V_k,
\]
where for each $i = 1, \dotsc, k$,
\begin{itemize}
\item[(i)] $V_i$ is a spherical space, or
\item[(ii)] $V_i$ is diffeomorphic to $S^2 \times S^1$, or
\item[(iii)] $V_i$ is aspherical (i.e., it's a $K(\pi,1)$ manifold).
\end{itemize}

If some $V_i$ is diffeomorphic to $S^2 \times S^1$, then $H_2(V_i) \neq 0$; hence $H_2(V) \neq 0$.

If some $V_i$ is aspherical, then by the positive resolution of the virtual positive first Betti number conjecture  (see \cite[Thm. 9.2]{Agol}\footnote{Specifically, a comment after Theorem 9.2 in \cite{Agol} points out that the theorem resolves question 17 in Thurston's influential list of 24 questions in \cite{Thurston}, which is, in fact, the virtual positive first Betti number conjecture. See also \cite{3_mfld_groups}.}), there is an $r$-fold cover $\tilde{V}_i$ of $V_i$ such that $\tilde{V}_i$ has positive first Betti number; hence $H_2(\tilde{V}_i) \neq 0$ by Poincar{\'e} duality. Let $X$ denote the three-manifold obtained by removing $V_i$ from the connected sum decomposition of $V$, i.e., $V = V_i \# X$. Simple cut and paste arguments show that
\[
\tilde{V} := \tilde{V}_i \# \underbrace{X \# \dotsb \#X}_{r\: {\rm times}} \:\:\:\: \text{ is an $r$-fold cover of } \:\:\:\: V_i \#X = V\,.
\]
Then $H_2(\tilde{V}) \neq 0$ since $H_2(\tilde{V}_i) \neq 0$.

\begin{figure}[t]
  \centering
\begin{tikzpicture}[scale=1]

\shade[color=black,ball color = blue, opacity = 0.3] (2,2) circle (1cm);
\shade[ball color = blue, opacity = 0.3] (-2,2) circle (1cm);
\draw [thick] (-2,2) circle (1cm);
\draw [thick] (2,2) circle (1cm);
\draw[dashed,thick] (-3,2) arc (180:0:1 and .2);
\draw[thick] (-3,2) arc (180:360:1 and .2);
\draw[dashed,thick] (1,2) arc (180:0:1 and .2);
\draw[thick] (1,2) arc (180:360:1 and .2);

\shadedraw[fill=white, opacity = .6]
(-1.5,2.7) to[out=90,in=180, looseness=.3]
(0,2.9) to[out=0,in=90, looseness=.3] (1.5,2.7) to[out=270,in=180, looseness=.7] (1.75,2.6) to[out=0,in=270, , looseness=.7] (2,2.7)
to[out=90,in=0, , looseness=.51] (0,3.2)
to[out=180,in=90, , looseness=.51] (-2,2.7)
to[out=270,in=180, , looseness=.7] (-1.75,2.6)
to[out=0,in=270, , looseness=.7] (-1.5,2.7);
\shadedraw[fill=blue, fill opacity=0.3,draw=black,
  draw opacity=1,thick, top color=blue!30!white]
(-1.5,2.7) to[out=90,in=180, looseness=.3]
(0,2.9) to[out=0,in=90, looseness=.3] (1.5,2.7) to[out=270,in=180, looseness=.7] (1.75,2.6) to[out=0,in=270, , looseness=.7] (2,2.7)
to[out=90,in=0, , looseness=.51] (0,3.2)
to[out=180,in=90, , looseness=.51] (-2,2.7)
to[out=270,in=180, , looseness=.7] (-1.75,2.6)
to[out=0,in=270, , looseness=.7] (-1.5,2.7);

\begin{scope}[yshift = 4cm, rotate = 180]
\shadedraw[fill=white, opacity = .6]
(-1.5,2.7) to[out=90,in=180, looseness=.3]
(0,2.9) to[out=0,in=90, looseness=.3] (1.5,2.7) to[out=270,in=180, looseness=.7] (1.75,2.6) to[out=0,in=270, , looseness=.7] (2,2.7)
to[out=90,in=0, , looseness=.51] (0,3.2)
to[out=180,in=90, , looseness=.51] (-2,2.7)
to[out=270,in=180, , looseness=.7] (-1.75,2.6)
to[out=0,in=270, , looseness=.7] (-1.5,2.7);

\shadedraw[fill=blue, fill opacity=0.3,draw=black,
  draw opacity=1,thick, top color=blue!30!white, shading angle=180]
(-1.5,2.7) to[out=90,in=180, looseness=.3]
(0,2.9) to[out=0,in=90, looseness=.3] (1.5,2.7) to[out=270,in=180, looseness=.7] (1.75,2.6) to[out=0,in=270, , looseness=.7] (2,2.7)
to[out=90,in=0, , looseness=.51] (0,3.2)
to[out=180,in=90, , looseness=.51] (-2,2.7)
to[out=270,in=180, , looseness=.7] (-1.75,2.6)
to[out=0,in=270, , looseness=.7] (-1.5,2.7);
\end{scope}

\draw [->,thick] (0,0.5) -- (0,0);

\shade[color=blue,ball color=blue,opacity=0.3] (-1,-1.5) arc (0:-180:1cm and 5mm) arc (180:0:1cm and 1cm);
\shade[ball color=blue,opacity=0.3] (3,-1.5) arc (0:-180:1cm and 5mm) arc (180:0:1cm and 1cm);
\draw[thick] (-3,-1.5) arc (180:0:1 and 1);
\draw[thick] (1,-1.5) arc (180:0:1 and 1);
\draw[dashed,thick] (-3,-1.5) arc (180:0:1 and .5);
\draw[thick] (-3,-1.5) arc (180:360:1 and .5);
\draw[dashed,thick] (1,-1.5) arc (180:0:1 and .5);
\draw[thick] (1,-1.5) arc (180:360:1 and .5);

\begin{scope}[yshift = -3.5cm]
\shadedraw[fill=white, opacity = .6]
(-1.5,2.7) to[out=90,in=180, looseness=.3]
(0,2.9) to[out=0,in=90, looseness=.3] (1.5,2.7) to[out=270,in=180, looseness=.7] (1.75,2.6) to[out=0,in=270, , looseness=.7] (2,2.7)
to[out=90,in=0, , looseness=.51] (0,3.2)
to[out=180,in=90, , looseness=.51] (-2,2.7)
to[out=270,in=180, , looseness=.7] (-1.75,2.6)
to[out=0,in=270, , looseness=.7] (-1.5,2.7);
\shadedraw[fill=blue, fill opacity=0.3,draw=black,
  draw opacity=1,thick, top color=blue!30!white]
(-1.5,2.7) to[out=90,in=180, looseness=.3]
(0,2.9) to[out=0,in=90, looseness=.3] (1.5,2.7) to[out=270,in=180, looseness=.7] (1.75,2.6) to[out=0,in=270, , looseness=.7] (2,2.7)
to[out=90,in=0, , looseness=.51] (0,3.2)
to[out=180,in=90, , looseness=.51] (-2,2.7)
to[out=270,in=180, , looseness=.7] (-1.75,2.6)
to[out=0,in=270, , looseness=.7] (-1.5,2.7);
\end{scope}

\draw (-4.25, -1.5) node [scale=.85]{$RP^3\# RP^3$};
\draw (-4.1, 2) node [scale=.85]{$S^1 \times \  S^2$};

\draw[white] (4.25, -1.5) node [scale=.85]{$RP^3\# RP^3$};

\end{tikzpicture}
\caption{\small The case $V = RP^3 \# RP^3$, which is double covered by $S^1 \times S^2$.}
\label{fig:RP3}
\end{figure}
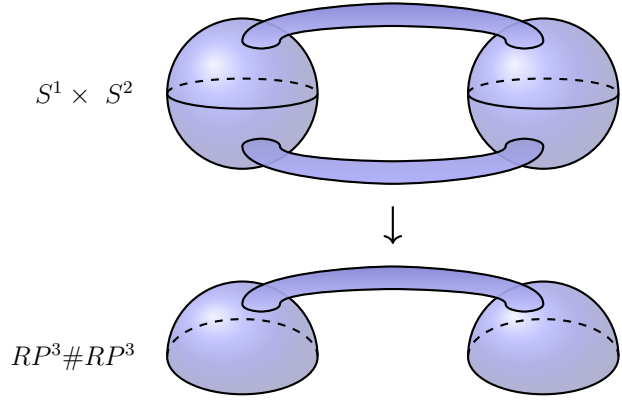

Lastly, suppose each $V_1, \dotsc, V_k$ is a nontrivial spherical space with $k \geq 2$. In this case, we can unravel $V_1$ to obtain a finite cover of the form $V_2 \# V_2 \# Y$ (see \cite{GL_topo_cosmo} for details). But $V_2 \# V_2$ is covered by two three-spheres connected by $s > 1$ handles, which has nontrivial second homology.  See Figure \ref{fig:RP3} for the special case $V =  RP^3 \# RP^3$, which is double covered by $\tilde{V} = S^1 \times S^2$, so that $H_2(\tilde{V}) \ne 0$.

\qed 

\medskip
\noindent
{\it Remark.}  The condition in Theorem \ref{thm: main} that $V$ is expanding in all directions is, of course, stronger than the mean expansion assumption in Hawking's cosmological singularity theorem.  The former condition can we weakened some using the notion of $k$-convexity.  It is sufficient to require that  the second fundamemental form $K$ be (strictly) $2$-convex, which means that the sum of the lowest two eigenvalues of $K$ is positive.  In this case, ${\rm tr}_\S K  > 0$ at each point of a surface $\S \subset V$, which is sufficient for our arguments.

\medskip

\section{A rigid version of Theorem \ref{thm: main}}

As mentioned in the introduction, in the recent paper \cite{GL_Penrose_rigid}, we studied rigidity aspects of Penrose's singularity theorem.  Specifically we address the following question: if a spacetime satisfies the hypotheses of Penrose's singularity theorem except that one assumes the existence of a weakly trapped surface, instead of a strictly trapped surface, then what can be said about the global structure of the spacetime if it is null geodesically complete?  General results  in  \cite{GL_Penrose_rigid} were used to obtain the following theorem in the cosmological setting (see \cite[Theorem 9]{GL_Penrose_rigid}).

%

\begin{thm}\label{thm: rigid 1}
Let $(M,g)$ be a spacetime that satisfies the null energy condition and is future null geodesically complete. Let $V$ be a smooth spacelike compact Cauchy hypersurface, and let $\S$ be a closed nonseparating two-sided hypersurface in $V$. Assume $\S$ is past weakly trapped.
Then either
\begin{itemize}
\item[\emph{(a)}] $M$ is past null geodesically incomplete, or else,
\item[\emph{(b)}] $M$ is foliated by two transverse families of closed totally geodesic null hypersurfaces; moreover, their intersections cover $M$ by totally geodesic codimension \emph{2} spacelike submanifolds.
\end{itemize}
\end{thm}

This theorem, together with our arguments involving the positive resolution of the virtual positive first Betti number conjecture, may be used easily to obtain the following rigid version of Theorem \ref{thm: main}, in which the second fundamental form is only assumed to be positive semi-definite.  


\begin{thm}\label{thm: rigid 2}
Let $(M,g)$ be a four-dimensional  spacetime that satisfies the null energy condition and is future null geodesically complete. Suppose $V$ is a smooth compact spacelike Cauchy surface such that its second fundamental $K$ is positive semi-definite.  Then one of the 
following holds:
\ben
\item [(i)] V is a spherical space.
\item [(ii)]  $M$ is past null geodesically incomplete. 
\item [(iii)]  $M$ (or a finite cover of $M$) is foliated by two transverse families of totally geodesic  null hypersurfaces; moreover, their intersections cover $M$ by totally geodesic two-dimensional spacelike submanifolds.
\een



%

\end{thm} 

\medskip
\noindent
\emph{Remarks.}

\medskip
\noindent
1. The Nariai spacetime, which is a product of two-dimensional de Sitter space and the two-sphere,  illustrates case (iii). The Cauchy surface for the Nariai spacetime is topologically $S^2 \times S^1$ which double covers $RP^3 \# RP^3$. The corresponding quotiented spacetime is an example of Theorem \ref{thm: rigid 2} where it is necessary to go to a finite cover.

\medskip
\noindent
2. Further results concerning rigidity aspects of Theorem \ref{thm: main} have recently been obtained by the second author along with C. Rossdeutscher, W. Simon and 
R. Steinbauer~(\cite{LRSS}). The results there could also be applied to the future-null-geodesically-complete setting of Theorem \ref{thm: rigid 2} with a similar conclusion (iii).



\section*{Acknowledgments} 
We thank the organizers---A. Shadi TahvildarZadeh, A. Burtscher, J. L. Flores, and L. Mehidi---of the meeting ``BIRS-IMAG meeting Geometry, Analysis, and Physics in Lorentzian Signature" for putting together a stimulating conference.  Eric Ling was supported by Carlsberg Foundation CF21-0680 and Danmarks Grundforskningsfond CPH-GEOTOP-DNRF151. We thank A. Piubello for designing the figures.

\vspace{.1in}
\noindent
{\bf Declarations.} On behalf of both authors, the corresponding author states that there is no conflict of interest. No data were collected or analyzed as part of this project.

\bibliographystyle{amsplain}

\end{document}